\input{aipcheck}

\documentclass[
    ,final            
  ]
  {aipproc}

\layoutstyle{6x9}

\begin{document}

\title{Activity of five WZ Sge-type systems in a few years after their
outbursts}


\author{E. Pavlenko}{
  address={Crimean Astrophysical Observatory, Crimea, Ukraine}
}
\author{O. Antoniuk}{
  address={Crimean Astrophysical Observatory, Crimea, Ukraine}
}
\author{K. Antoniuk}{
  address={Crimean Astrophysical Observatory, Crimea, Ukraine}
}
\author{M. Andreev}{
  address={Terskol Branch of the RAS Institute of Astronomy, Russia}
}
\author{D. Samsonov}{
  address={Crimean Astrophysical Observatory, Crimea, Ukraine}
}
\author{A. Baklanov}{
  address={Crimean Astrophysical Observatory, Crimea, Ukraine}
}
\author{A. Golovin}{
  address={Main Astronomical Observatory of the National Academy of Sciences, Ukraine}
}
\begin{abstract}
We present results of observations and analysis of five WZ-type
dwarf novae: SDSS J080434.20+510349.2, SDSS J102146.44+234926.3,
V1108 Her, ASAS J0025 and WZ Sge in 3 - 7 years after their
outbursts. We found: 1) the observational evidences that emerging
of the white dwarf nonradial pulsation  in SDSS
J080434.20+510349.2 [1] is no longer a regular since 3-rd year
after 2006 outburst; 2) the true value of orbital period for V1108
Her (0.05672 d) and mass ratio (0.09); 3) The main feature of
quiescent light curves of ASAS J0025 is the high-amplitude
quasi-periodical light variations probably originated in accretion
disk; 4) in SDSS J102146.44+234926.3 in some occasions the
brightness is modulated by period of 0.01367 day or with the
multiple of it.

The main brightness modulation of J080434.20+510349.2, V1108 Her
and WZ Sge is the orbital one and has two-humped shape.

\end{abstract}

\maketitle

\section{introduction}
WZ Sge-type stars are the close binary systems at a latest stage
of evolution. They consist from a white dwarf (WD) accreting
matter from a late (probably brown dwarf) type secondary. These
binaries are attractive objects for search for the white dwarf
pulsations. During outburst the observed brightness variability of
WZ Sge type stars is caused mostly by the thermal and tidal disk
instability. The orbital modulation also could contribute in
high-inclined systems during this state. The probability to detect
of the WD pulsations depends on the quantity of sources of
variable radiation and on the temperature of WD at the time of
observation. The quiescent state seems to be more preferable for
this task because of reduced number of sources of variable
radiation. However it is difficulty to predict when the white
dwarf should pass its instability strip after the outburst for
each individual case.

We conducted the observations of five WD Sge-type dwarf novae
J080434.20+510349.2, SDSS J102146.44+234926.3, V1108 Her, ASAS
J0025 and WZ Sge itself in 3--7 years after their outbursts and
present the analysis of activity of these systems.

\section{Observations}

CCD observations of the five  WZ Sge-type stars were carried out
using different telescopes of the Crimean astrophysical
observatory (Ukraine) and Terskol observatory (Russia). The
observations were made in $R$ spectral band or in unfiltered
light. The log of observations is given in Table 1, where the site
of observations, telescopes, date of the outburst preceding
observation and date of observations are presented.

\section{SDSS J080434.20+510349.2}

This object contains the accreting WD pulsator. The first
appearance of 12.6-min pulsations were detected in 8-9 month after
the 2006 outburst [1]. We continued a monitoring of SDSS
J080434.20+510349.2 (hereafter, SDSS J0804) in 2009. The original
data folded on the orbital period 0.0590048 d [2] are shown in
Fig. 1 (left). The orbital profile is a two-humped one, sometimes
one of humps could be more or less  divided  into two (JD 2454852
and JD 2454882). It is seen that in some occasions the orbital
modulation is slightly affected by another short-term variability.
Periodograms calculated for these data (using ISDA package [3])
are shown in Fig. 2 (right). Besides of peaks pointing to the
orbital period (Porb) and its half value Porb/2, the periodograms
for only two from six rows of observations (JD 244852 and JD
244882) contain peak at period 740 s (12.33 min) and it's twice
value. In the limits of width of peak it coincides with the known
period of pulsation 12.6 min. So in $\sim$ 3 year after the 2006
outburst the appearance of pulsations became irregular. It is
worth to note that the complex structure of the orbital light
curve produce additional peak on the light curve at Porb/4. More
of a linear combinations of the orbital and pulsation frequencies
appear for these data.
\begin{table}
\begin{tabular}{lrrrr}
\hline
  & \tablehead{1}{r}{b}{Site}
  & \tablehead{1}{r}{b}{Year of\\
outburst}
  & \tablehead{1}{r}{b}{Year of  \\observations}\\
\hline
SDSS J080434.20+510349.2 & CrAO (2.6-m), Terskol (2-m) & 2006 & 2009\\
V1108 Her & CrAO (2.6-m) & 2004 & 2008\\
SDSS J102146.44+234926.3 & CrAO (2.6-m) & 2006 & 2009\\
ASAS J0025 & Terskol (60 cm) & 2004 & 2007\\
WZ Sge & CrAO (2.6-m, 1.25-m) & 2001 & 2008\\
\hline
\end{tabular}
\caption{Log of observations for five WZ Sge-type stars}
\label{tab:a}
\end{table}

\begin{figure}
  \includegraphics[height=0.7\textheight]{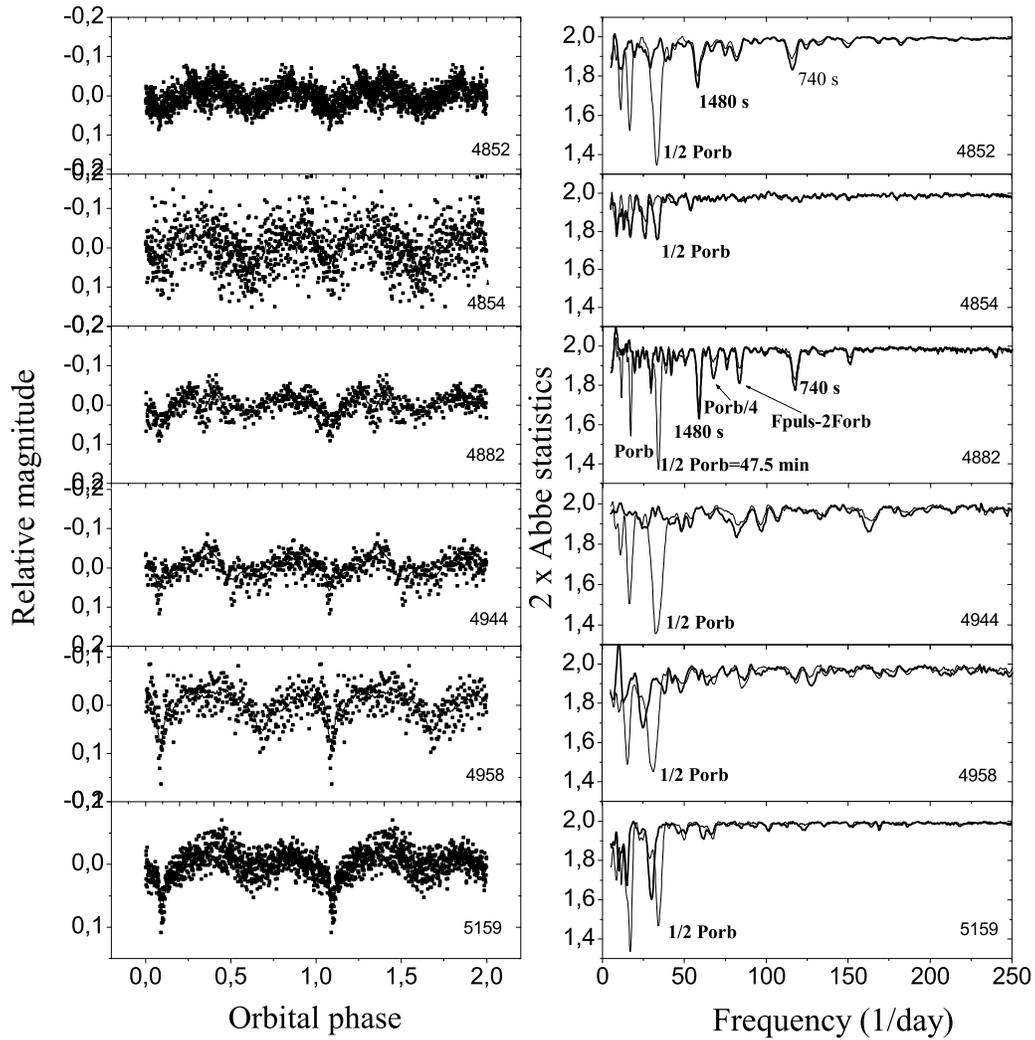}
  \caption{Left: original data of SDSS J0804 folded on the orbital period of 0.0590048 day. Right:
  corresponding periodograms. The periodograms for original data are designed by thin line while
  these for data after orbital period subtraction, by thick line}
\end{figure}

\section{V1108 Her}

The last known measuring of the orbital period of V1108 Her was
done by Price et al. [4] in a few days after the end of the 2004
outburst. Accordingly to their observations, period was
0.056855(69) d. We obtained our observation in 2008, believing
that in 4 years after the 2004 outburst the data may not be
affected by positive superhumps. The periodogram for the data
covering 8 nights of observations is shown in Fig. 2 (left). The
strongest peak points to period 0.05672(4) d. The corresponded
fractional period excess [5] (for period of superhump 0.057780 d
[4]) is 0.0187, so the mass ratio of components is 0.09. This
confirms the previous suggestion of Price et al. [4] that V1108
Her could be the "period bouncer". The data folded on the orbital
period are shown in Fig. 2 (right). One could see that the main
brightness variations are caused by two-humped orbital variations
with unequal mean amplitudes of neighbor humps ($0^{m}.05$ and
$0^{m}.025$). The shape of the orbital light curve resembles these
of high-inclined WZ Sge-type systems.
\begin{figure}
  \includegraphics[height=0.4\textheight]{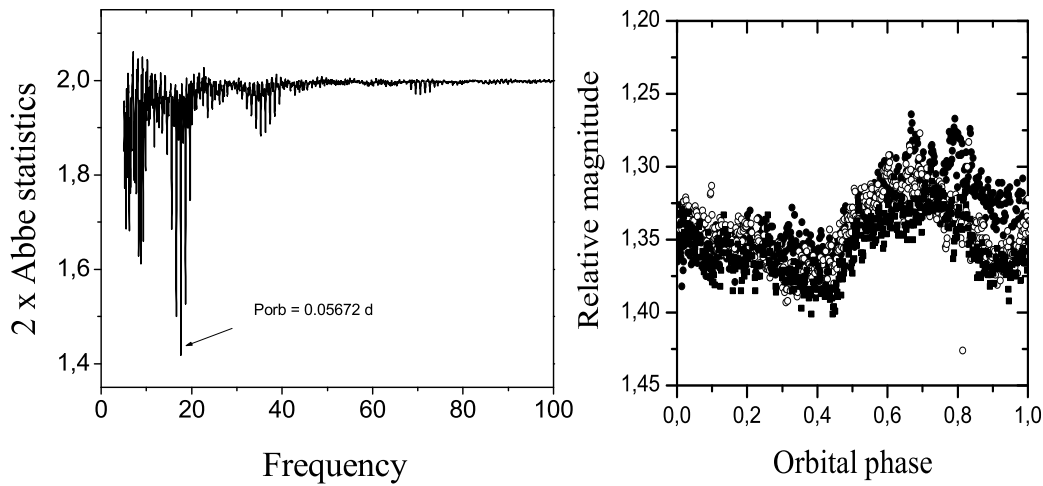}
  \caption{Left: periodogram of V1108 Her for 4 nights of observations in 2008.
  Frequency is expressed in  units of 1/day. Right: data folded on the orbital period}
\end{figure}

\section{SDSS J102146.44+234926.3, ASAS J0025 and WZ Sge}

The true orbital period of SDSS J102146.44+234926.3 is unknown.
Our observations in 2009 did not reveal expected orbital
modulation. Instead for some nights we detected the brightness
variations with 20.6 and 20 min period (Fig. 3) or with the
multiple of them. The mean amplitude for Jan 20/21, 2009 was
$0^{m}.1$. The source of this irregular periodicity is unclear.
Potentially it could be connected with instability of outer parts
of accretion disk, or with WD pulsations, or equal to the 1/4 of
expected value of orbital period.

ASAS J0025 displays a high-amplitude nightly variability with
amplitude up to $0^{m}.3$ (Fig. 4). However the data obtained in
2007 can't confirm the 0.05654 d candidates to the orbital period
suggested by Templeton et al. [6]. Instead a prominent variability
in a vicinity of suggested orbital period was observed (Fig. 5).
Such behavior resembles the variability often observing in a
low-inclined systems (for example in MV Lyr [7]) connected
probably with instability in accretion disk.

WZ Sge is eclipsed system. In 7 years after 2001 outburst the main
brightness variations are caused by orbital modulation with period
of 81-min  [8]. The example of two-humped light curve is presented
in Fig. 6. The amplitude of humps varies from cycle to cycle in a
region of $0^{m}.15-0^{m}.3$. There are variations of much smaller
amplitude superimposing the orbital light curves.

\begin{figure}
  \includegraphics[height=0.36\textheight]{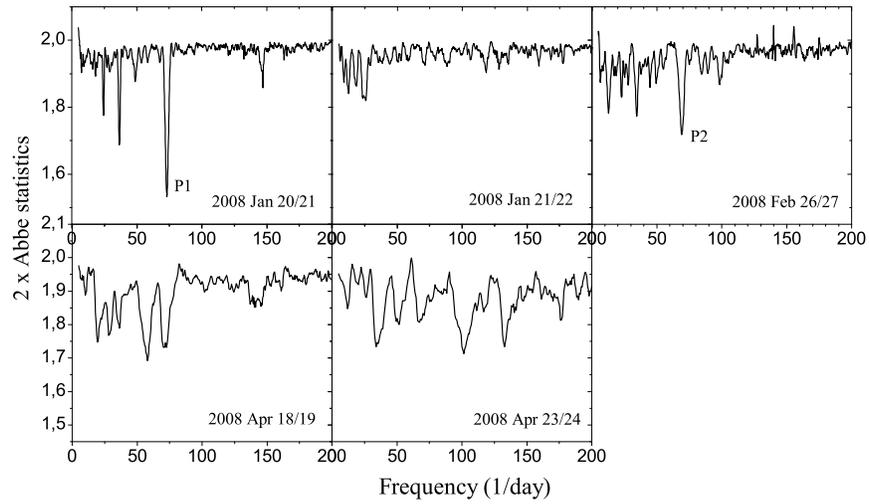}
  \caption{Periodogram of SDSS J102146.44+234926.3 for different data in 2009}
\end{figure}

\begin{figure}
  \includegraphics[height=0.2\textheight]{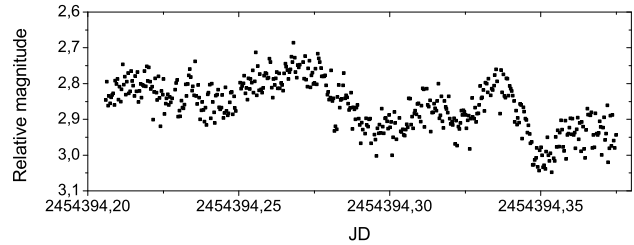}
  \caption{The example of nightly light curve of ASAS J0025 in 2007}
\end{figure}

\begin{figure}
  \includegraphics[height=0.25\textheight]{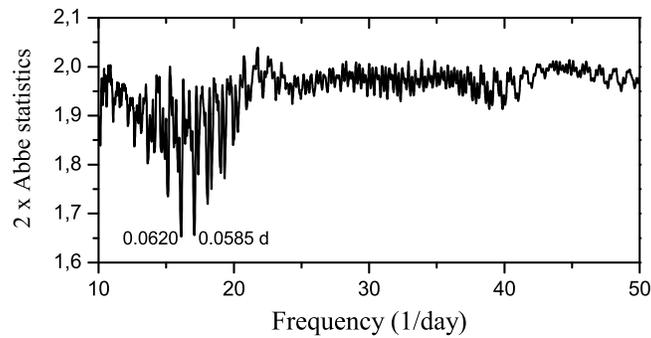}
  \caption{The periodogram of ASAS J0025 for 4 nights in 2007}
\end{figure}

\begin{figure}
  \includegraphics[height=0.3\textheight]{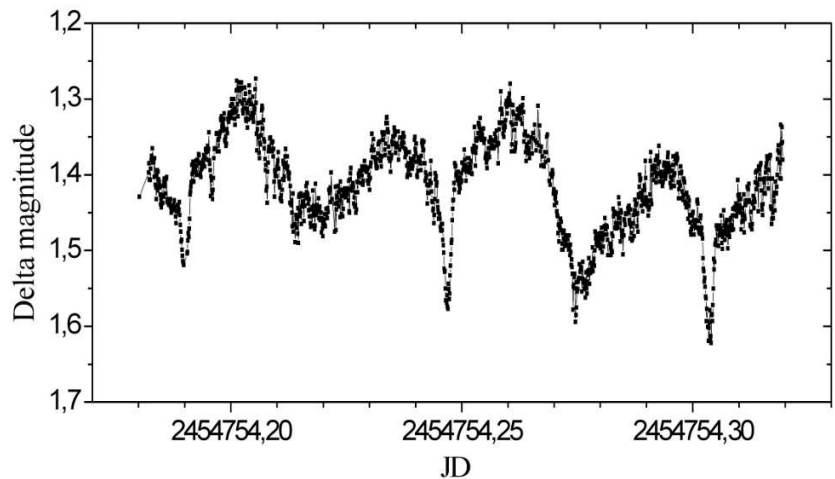}
  \caption{Example of the WZ Sge light curve in 2008}
\end{figure}

\section{Conclusion}

We considered the behavior of five WZ Sge-type stars in several
years after their outburst. The main brightness variations of the
high-inclined systems (WZ Sge, SDSS J0804, V1108 Her) is caused by
the orbital modulation. In SDSS J0804 in some occasions the WD
pulsations superimpose the orbital light curves. SDSS
J102146.44+234926.3 and ASAS J0025 display brightness variability
which nature is unclear.

\begin{theacknowledgments}
  E. Pavlenko and A. Golovin are grateful to the LOC for possibility to
  participate in the conference.
  This work was partially supported by grant 28.2/081 of
  Ukrainian Fund of Fundamental Researches.
\end{theacknowledgments}



\bibliographystyle{aipproc}   

\bibliography{sample}

\IfFileExists{\jobname.bbl}{}
 {\typeout{}
  \typeout{******************************************}
  \typeout{** Please run "bibtex \jobname" to optain}
  \typeout{** the bibliography and then re-run LaTeX}
  \typeout{** twice to fix the references!}
  \typeout{******************************************}
  \typeout{}
 }

\end{document}